\documentclass[
reprint,
superscriptaddress,
amsmath,
amssymb,
aps,
prl,
twocloumn]{revtex4-2}
\usepackage{graphicx}
\usepackage{dcolumn}
\usepackage{bm}
\usepackage{physics}
\usepackage{appendix}
\usepackage[normalem]{ulem}
\usepackage{comment}
\usepackage{subfigure}
\usepackage{svg}
\usepackage{todonotes}
\usepackage{csquotes}
\usepackage{bbm}
\usepackage{dsfont}
\usepackage{hyperref}
\usepackage[capitalise]{cleveref}
\usepackage{placeins}
\newtheorem{lemma}{Lemma}

\usepackage{xcolor}
\begin{document}

\preprint{APS/123-QED}

\title{Spectral Gap Optimization for Enhanced Adiabatic State Preparation}

\author{Kshiti Sneh Rai}
\email{ksrai@lorentz.leidenuniv.nl}
\affiliation{Instituut-Lorentz, Universiteit Leiden, P.O. Box 9506, 2300 RA Leiden, The Netherlands}
\affiliation{$\langle aQa^L\rangle$ Applied Quantum Algorithms Leiden, The Netherlands}
\author{Jin-Fu Chen}
\email{jinfuchen@lorentz.leidenuniv.nl}
\affiliation{Instituut-Lorentz, Universiteit Leiden, P.O. Box 9506, 2300 RA Leiden, The Netherlands}
\affiliation{$\langle aQa^L\rangle$ Applied Quantum Algorithms Leiden, The Netherlands}
\author{Patrick Emonts}%
\email{emonts@lorentz.leidenuniv.nl}
\affiliation{Instituut-Lorentz, Universiteit Leiden, P.O. Box 9506, 2300 RA Leiden, The Netherlands}
\affiliation{$\langle aQa^L\rangle$ Applied Quantum Algorithms Leiden, The Netherlands}
\author{Jordi Tura}%
\affiliation{Instituut-Lorentz, Universiteit Leiden, P.O. Box 9506, 2300 RA Leiden, The Netherlands}
\affiliation{$\langle aQa^L\rangle$ Applied Quantum Algorithms Leiden, The Netherlands}

\date{\today}
\begin{abstract}
The preparation of non-trivial states is crucial to the study of quantum many-body physics.
Such states can be prepared with adiabatic quantum algorithms, which are restricted by the minimum spectral gap along the path.
In this letter, we propose an efficient method to adiabatically prepare tensor networks states (TNSs).
We maximize the spectral gap leveraging degrees of freedom in the parent Hamiltonian construction. 
We demonstrate this efficient adiabatic algorithm for preparing TNS, through examples of random TNS in one dimension, AKLT, and GHZ states.
The Hamiltonian optimization applies to both injective and non-injective tensors, in the latter case by exploiting symmetries present in the tensors. 
\end{abstract}

\maketitle
\emph{\label{sec:level1}Introduction.---}
The preparation of an initial state of interest is the basis for most quantum experiments~\cite{bernien_probing_2017,ebadi_quantum_2021,barends_digitized_2016}. 
Given its importance and inherent challenges, quantum many-body state preparation has been approached from various directions~\cite{montanaro_quantum_2016}.
These include gate-based methods~\cite{schon_sequential_2005,ran_encoding_2020,wei_efficient_2023,fomichev_initial_2023,malz_preparation_2024} which are implementable on digital architectures~\cite{wendin_quantum_2017,barends_digitized_2016}.
However, the task is not limited to digital devices, as it also constitutes a native operation on analog quantum simulators~\cite{cirac_goals_2012,gross_quantum_2017}.

Adiabatic state preparation involves interpolating a system from the ground state of a trivial Hamiltonian to that of a final Hamiltonian.
One key factor influencing the performance of adiabatic algorithms is the minimal spectral gap $\Delta$ during the interpolation. 
Rigorous estimates on the runtime to prepare the ground state of a Hamiltonian yield a scaling $\propto 1/\Delta^3$~\cite{jansen_bounds_2007}.
Thus, developing methods to increase  $\Delta$ is desirable for more efficient preparation algorithms~
\cite{cruz_preparation_2022,wei_efficient_2023,zhu_efficient_2024}.

Tensor networks (TNs) comprise a vast of family of quantum many-body states that offer additional control over $\Delta$ via the so-called parent Hamiltonian (PH) construction.
TNs are characterized by an area-law scaling of entanglement across bipartitions~\cite{verstraete_matrix_2006}, and can be expressed concisely with polynomially many parameters (in the system size), as opposed to exponential number of parameters in the general case. 
Moreover, they provide efficient polynomial approximations of ground states of 1D gapped local Hamiltonians~\cite{cirac_matrix_2021,hastings_spectral_2006,hastings_area_2007}.
Conversely, TNs serve as ground states of a family of local PH, which can be constructed using the tensor representations~\cite{fannes_finitely_1992,schuch_classifying_2011,perez-garcia_matrix_2007}.
An important mathematical property of the tensors is injectivity, which describes the relation between the boundary and bulk of a TN state. 
If a state has an injective TN representation, then it enables the construction of PHs with \textit{unique} ground states~\cite{perez-garcia_peps_2007}, even in arbitrary lattices. 
Examples of injective matrix product states (MPS), which are TNs in 1D, include the AKLT state and the cluster state.

The flexibility in the space of PHs allows for the optimization of various quantities. 
For example in Ref.~\cite{giudici_locality_2022}, the proposed algorithm searches for the PH which has an efficient approximation in terms of a Hamiltonian of smaller locality. 
In this work, we actively exploit the non-uniqueness of the PH to optimize the gap during an adiabatic evolution.
The algorithm is comprised of the following two steps: 
(i) We build the parametrized family of PHs with a fixed locality, of a \textit{single} TN state. 
Following \textit{Weyl's inequality}~\cite{weyl_asymptotische_1912}, the spectral gap $\Delta$ is a concave function of the free parameters in the PH.
The concavity implies that the maximum gap $\Delta_{\mathrm{opt}}$ and corresponding optimal parameter values of the Hamiltonian can be obtained efficiently using \textit{e.g.} a gradient-ascent algorithm.
(ii) We fix a deformation of tensors $A(\lambda)$ which  interpolates from a product state $(\lambda=0)$ to a state of interest $(\lambda=1)$, ensuring that the intermediate states are also TNs.
The TN formulation along the path allows pointwise implementation of the gap optimization, finally providing a Hamiltonian path $H_{\mathrm{opt}}(\lambda)$ which has maximal gap at all intermediate points. 

We illustrate our gap optimization algorithm on exemplary classes of states: random injective MPS, the AKLT state, and the non-injective GHZ state.  
In addition, our machinery can effectively incorporate symmetries which allow the method's scalability to larger system sizes and to circumvent the closing of the gap in systems with degenerate ground states. 
Furthermore, we derive an analytical description of the optimal parameter evolution as a differential equation, which can replace the need for gradient-ascent optimization.
The techniques can also be applied when the tensor representation of a TN state is unknown, but one of its PHs is given. 

Our work contributes to efficient adiabatic preparation methods for TNSs. These are vital in quantum information processing, as they enable protocols for initializing states in quantum phase estimation~\cite{schon_sequential_2005, ran_encoding_2020, fomichev_initial_2023} and facilitate the study of phenomena such as quantum quenches, dynamics, and appearance of thermalization in isolated quantum systems~\cite{polkovnikov_colloquium_2011}. 
Moreover, the utility of adiabatic methods extends to quantum device verification, as highlighted in~\cite{cruz_preparation_2022}.
\begin{figure}
    \centering
    \includegraphics[width=\linewidth]{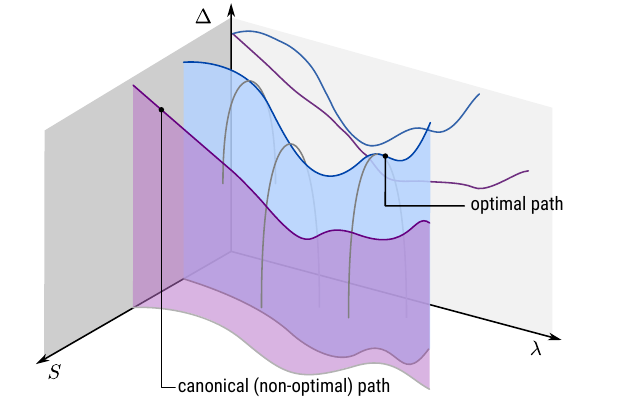}
    \caption{
    Sketch of the optimization algorithm.
    The blue path is the optimized path through the space of Hamiltonians, i.e. it has the maximal gap.
    The convex optimization in the parameters $\boldsymbol{S}$ is illustrated by grey lines.
    For comparison, the purple curve is an initial, non-optimal guess for a path.
    The actual gap along the path is visualized as a projection on the light-grey $\Delta-\lambda$ plane.}
    \label{fig:Sketch_algorithm}
\end{figure}

\emph{Preliminaries.---}
Tensor networks (TNs) constitute a class of states that encode the physical state in a set of tensors which are placed on the vertices of a graph.
If two vertices are connected, the associated tensor dimensions are contracted.
Concretely, in one spatial dimension, an MPS with periodic boundary conditions can be written as
\begin{align}
\label{eq:mps_def}
    \ket{\psi}= \sum_{\{\sigma\}}\sum_{\{i\}}A^{(0),\sigma_0}_{i_{0},i_1}A^{(1),\sigma_1}_{i_1,i_2}\dots A^{(N-1),\sigma_{N-1}}_{i_{N-1},i_0}\ket{\sigma_0\dots\sigma_{N-1}}, 
\end{align}
where $A^{(i),\sigma}_{\alpha,\beta}$ is the tensor on site $i$, $\sigma$ is its physical index ($d$-dimensional) and the virtual indices $\alpha,\beta$ ($D$-dimensional) are used for the contraction (c.f.~\cite{suppmat}).
Given a TNS, we can construct the \emph{local} PHs, i.e. a family of Hamiltonians such that the TNS is the zero energy ground state of each of them.
In general, it is trivial to generate a PH for any state by setting $H=\mathds{1}-\ket{\psi}\bra{\psi}$.
However, this Hamiltonian is not local by construction.
For TNs, the PH $H=\sum_{i=1}^N h_i$ of an $N$-site system can be constructed with the local term
\begin{align}
\label{eq:local_parent_hamiltonian}
    h_i = \sum_{\alpha,\beta=1}^M S_{\alpha\beta}\ket{\phi_{i,\alpha}}\bra{\phi_{i,\beta}}=\Phi_i \bm{S} \Phi_i^\dagger,
\end{align}
where $\bm{S}$ is an arbitrary positive definite matrix normalized to $\mathrm{tr}(\bm{S})=1$, and $\Phi_i=(\ket{\phi_{i,1}},\ket{\phi_{i,2}},...,\ket{\phi_{i,M}})$ is a basis depending on the tensors of the initial state.
For constructing the basis $\Phi_i$, the tensor needs to be \textit{injective}, which is achieved by blocking $L$ tensors together, such that $d^L\geq D^2$.
This makes the map from the virtual indices to physical indices injective, which is required to build PHs with unique ground states~\cite{perez-garcia_peps_2007}.
In the literature, the \textit{canonical} choice of a PH is taken to be $\bm{S} = \mathbbm{1}/M$.
However, the freedom in $\bm{S}$ can be used to optimize certain properties of the PH, e.g. its locality~\cite{giudici_locality_2022} or its gap. 
In this work we consider the latter.

\emph{Adiabatic path optimization.---} 
Adiabatic algorithms are originally formulated as an interpolation between Hamiltonians as $H = (1-s)H_0+sH_1$, where $s\in [0,1]$ is the control parameter.
However, in this letter, we choose a different perspective by deforming the tensors representing the initial and final states.
Given an efficient TN representation of the final state, one can express the evolution of the tensor $A:= A(\lambda)$ from a $\ket{\psi_{\text{in}}}$ to $\ket{\psi_{\text{fi}}}$, where each tensor entry is a continuous function of parameter $\lambda$.
This interpolation is chosen such that $\ket{\psi_{\text{in}}}$ is given by $A(0)$ and $\ket{\psi_{\text{fi}}}$ by $A(1)$.
As a result of this parametrization, the state is represented by a TN at all intermediate points, allowing the construction of PHs along the path.
Ultimately, this makes it viable to optimize the adiabatic path taken through the Hamiltonian space.
We remark, contrary to the usual approaches~\cite{schiffer_adiabatic_2022,negretti_speeding_2013,chen_speeding_2022}, here we do \emph{not} optimize the speed or scheduling of the parameter $s$.
Since the Hamiltonian optimization is distinct from scheduling tasks, it is possible to combine both for improving the overall performance of adiabatic algorithms.

Knowing the initial and the final tensors, we interpolate between them with $A^{(i),\sigma}_{\alpha,\beta}(\lambda)$ and for each value of $\lambda$, we optimize the gap of the PH.
For every $\lambda$, the input of the optimization procedure is comprised of the MPS formulation of the state given by $A(\lambda)$ as in \cref{eq:mps_def}. 
The concavity of the gap function (c.f.~\cite{suppmat}) enables efficient convergence through a gradient descent algorithm along the parameters $\boldsymbol{S}$.
First, we obtain $\frac{\partial H}{\partial \boldsymbol{S}}$ numerically, and then compute the expectation values of this derivative in the excited state.
For small systems, this can be achieved by an exact diagonalization procedure.
However, as the size of the system increases, finding low-energy eigenstates and the corresponding spectral gap - becomes computationally intensive~\cite{cubitt_undecidability_2015}.
In these cases, we employ excited state density matrix renormalization group (DMRG) to compute both the ground state and the first excited state, i.e. the gap~\cite{white_density_1992,schollwock_density-matrix_2011,stoudenmire_studying_2012}.
Performing the optimization consistently yields higher ground state preparation efficiencies and shorter runtimes, although the extent of improvement varies depending on the specific example. 

The numerical framework presented above suggests the existence of an ideal deformation path, for a given MPS interpolation $\Phi(\lambda)$.
We define the \textit{condition of optimality}, which ensures that the gap reaches an extremum at all values of $\lambda$ during the evolution.
Optimality implies that the gradients with respect to the elements of the matrix $\bm{S}$ are zero at the optimal point,
\begin{align}
\label{eq:condition_of_optimality}
    \frac{\partial\Delta(\Phi(\lambda),\bm{S})}{\partial\bm{S}}\bigg\vert_{\bm{S}=\bm{S}_{\mathrm{opt}}(\lambda)} = 0,\quad\forall\,\lambda\in[0,1],
\end{align}
where $\bm{S}_{\mathrm{opt}}(\lambda)$ denotes the optimal positive definite matrix (c.f. \cref{eq:local_parent_hamiltonian}).
An equivalent formulation of \cref{eq:condition_of_optimality} can be given in terms of the first excited state $\ket{\Psi_1}$ and $\Phi$ using $\chi$ defined as follows,
\begin{align}
    \chi_{\alpha\beta} := \sum_i \bra{\Psi_{1}}\ket{\phi_{i,\alpha}} \bra{ \phi_{i,\beta}}\ket{\Psi_{1}}.\label{eq:chi_alphabeta_definition}
\end{align}
At the optimal point, under the assumption that $\chi$ is full-rank, $\bm{S}_\mathrm{opt}$ and $\bm{\chi}_\mathrm{opt}$ are simultaneously diagonalizable, and additionally, all eigenvalues of $\bm{\chi}_\mathrm{opt}$ are equal to $\Delta_\mathrm{opt}$, i.e.,
$\bm{\chi}_\mathrm{opt}=\Delta_\mathrm{opt}\mathbbm{1}$,
where $\Delta_\mathrm{opt}=\Delta(\Phi ,\bm{S}_\mathrm{opt})$ is the optimal gap.
Generically, $\bm{S}_\mathrm{opt}$ is full rank and the gap is smooth at $\bm{S}_\mathrm{opt}$  although $\bm{S}_\mathrm{opt}(\lambda)$ might not be smooth be itself as we will show in examples.

Using the condition of optimality, it is possible to derive an ordinary differential equation (ODE) describing the evolution of $\bm{S}_\mathrm{opt}$ with changing $\lambda$. 
Differentiating \cref{eq:condition_of_optimality} with respect to $\lambda$ on both sides yields the following ODE:
\begin{align}
\label{eq:ODE_optimal_path}
    \partial_\lambda\bm{S}_\mathrm{opt}
    =
    -\left(\bm{\mathcal{H}}_{\Delta,\bm{S}_\mathrm{opt}}\right)^{-1}
    \partial_\Phi\bm{\mathcal{J}}_{\Delta,\bm{S}_\mathrm{opt}}\cdot\partial_{\lambda}\Phi,
\end{align}
where $\bm{\mathcal{J}}_{\Delta,\bm{S}_\mathrm{opt}(\lambda)}=\partial_{\bm{S}} \Delta|_{\bm{S}=\bm{S}_\mathrm{opt}(\lambda)}$ and $\bm{\mathcal{H}}_{\Delta,\bm{S}_\mathrm{opt}(\lambda)}=\partial^2_{\bm{S}}\Delta|_{\bm{S}=\bm{S}_\mathrm{opt}(\lambda)}$ represent the Jacobian and Hessian of the gap $\Delta$ at the optimal $\bm{S}_\mathrm{opt}(\lambda)$ (c.f.~\cite{suppmat}), respectively.
 This formulation leverages optimality (\cref{eq:condition_of_optimality}), bypassing the need for gradient-based optimization and enabling a generalization to higher-dimensional PEPS. Moreover, since it takes the first excited state as input, protocols for eigenstate preparation could benefit from this formulation~\cite{santagati_witnessing_2018,tilly_variational_2022}.

In the remainder of this letter, we will illustrate the gap optimization. 
We consider three examples based on interpolation from a product state to (i) a random injective MPS~\cite{cruz_preparation_2022}, (ii) an AKLT state~\cite{affleck_rigorous_1987} and (iii) a non-injective GHZ state. 

\begin{figure}[t!]
\centering
 \includegraphics[width=\columnwidth]{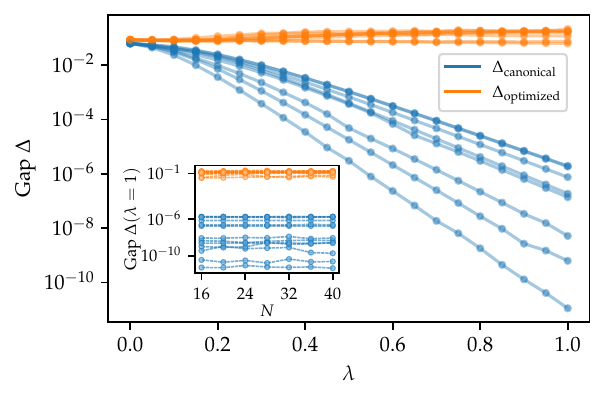}
\caption{Gap optimization for the preparation of many random injective MPS on $N=24$ sites.
In all cases, the gap along the path changes from exponentially decaying ($\Delta_{\mathrm{canonical}}$) to roughly constant ($\Delta_{\mathrm{optimized}}$). 
In the inset, we show the improvement in the gap at $\lambda=1$ for random seeds and system sizes from $N=16$ to $N=40$. } 
\label{fig:random_mps_path}
\end{figure}

\emph{Random injective MPS preparation.---}
First, we consider the preparation of 
a random injective MPS using a method proposed by Cruz \emph{et~al}~\cite{cruz_preparation_2022}.
A matrix product state on a chain of $d-$dimensional spins is obtained by applying a transformation to a product state of maximally entangled pairs,
\begin{align}
    \ket{\psi(\mathcal{P})} = \mathcal{P}^{\otimes N}\left(\sum_{i=1}^d\ket{ii}\right)^{\otimes N},
\end{align}
where $\mathcal{P}$ is defined as a map from the virtual to physical indices. 
As illustrated in Ref.~\cite{cruz_preparation_2022}, a smooth path of PHs can be generated by decomposing the map $\mathcal{P}$.
With a polar decomposition $\mathcal{P} = QW$, we obtain an isometry $W$ and a positive definite matrix $Q$.
In the case of injective MPS with $d^L = D^2$, the matrix $W$ is a unitary.
By parametrizing $Q = e^{\lambda K_1}$ and $W=e^{iK_2t}$, where $K_1$ and $K_2$ are random Hermitian matrices, it is possible to prepare a set of random injective MPS.
In Ref.~\cite{cruz_preparation_2022}, a canonical form for the PH $H(\lambda,t)$ based on the tensor deformation is proposed which we specialize for translation-invariant MPS (c.f.~\cite{suppmat}).
Here, we generate the MPS with random $(Q,W)$ on a $N=24$ sites spin-$\frac{1}{2}$ system with bond dimension $D=2$, and perform the Hamiltonian optimization procedure along the tensor deformation path. 
\Cref{fig:random_mps_path} compares the spectral gap of canonical Hamiltonian $H(\lambda,t)$ in comparison with the optimized Hamiltonian $H_{\mathrm{opt}}(\lambda,t)$.
$\Delta_{\mathrm{canonical}}(\lambda,t)$ is found to decay exponentially with $\lambda$, while after the optimization, the gaps of $\Delta_{\mathrm{optimized}}(\lambda,t)$ are approximately constant during the evolution. 
In the inset, we show that the difference between the optimized and unoptimized gaps of the PH of the final MPS remains exponentially large with increasing system size $N$.

\emph{Adiabatic preparation of AKLT state. ---}
Next, we optimize the preparation of the 1D AKLT state starting from a trivial product state~\cite{affleck_rigorous_1987}. 
In contrast to random MPS states, this state has considerably more symmetries which we will include later.
We interpolate the MPS tensors along a linear path,
\begin{align}
\label{eq:aklt_path_mps}
    A^{(-1)}=\begin{pmatrix}
        0&0
        \\
        \lambda&0
    \end{pmatrix},
    A^{(0)}=\frac{1}{\sqrt{2}}\begin{pmatrix}
    1&0
        \\
        0&-\lambda
    \end{pmatrix},
    A^{(+1)}=\begin{pmatrix}
        0&-\lambda
        \\
        0&0
    \end{pmatrix}
\end{align}
This MPS becomes injective upon blocking two sites, and the dimension of the kernel of the tensor map is $d^2-D^2 = 5$.
The basis states $\{\ket{\phi_i(\lambda)}\}_{i=1}^5$ for the kernel are parameterized such that at $\lambda=1$ they represent the Dicke states of the spin-2 subspace. 

\begin{figure}[t!]
    \centering\includegraphics[width=\columnwidth]{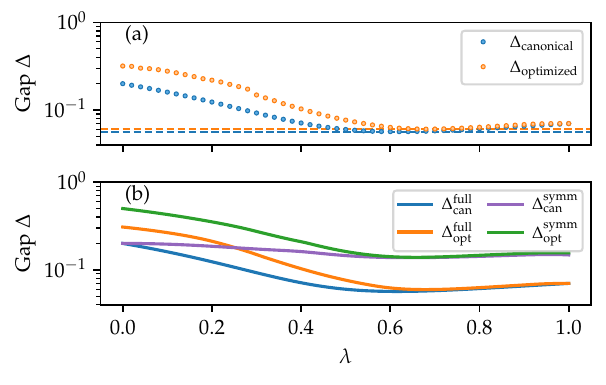}
    \caption{
        Depiction of AKLT state preparation. 
        (a) $N=30$ sites: the blue dots show gap of canonical Hamiltonian with $S_{11}=S_{22}=0.2$ in Eq.~\eqref{eq:aklt_S_mat} and orange dots show the gap after intermediate optimization. Dashed lines indicate the minimum gap along each path.
        (b) $N=8$ sites: preserving symmetries enhances the gap. The labels \textit{full} and \textit{symm} refer to the full Hilbert space and symmetric subspace with the ground state respectively. $\Delta_{\mathrm{can}}$ and $\Delta_{\mathrm{opt}}$ denote canonical and optimized gaps respectively.
        } 
    \label{fig:aklt_mps_path}
\end{figure}

In many physical cases, the spectral gap can be enhanced additionally by exploiting the symmetry of the MPS tensors.
The symmetries~\cite{moudgalya_exact_2018} in the AKLT state include (i) translation invariance, since it is a periodic MPS,  (ii) conservation of total $S^z$: $\expval{S^z_{\mathrm{tot}}}=0$, which appears due to the valence bond structure of the state and
(iii) a type of $\mathbb{Z}_2$ symmetry: with $Q=\prod_{i=1}^N S_i^x R$, where the operation $R$ reverses the ordering of the spins. 
Upon including these symmetries, the positive matrix $\boldsymbol{S}$ simplifies to 
\begin{align}
\label{eq:aklt_S_mat}
    \bm{S}=\mathrm{diag}(S_1,S_2,1-2S_1-2S_2,S_2,S_1).
\end{align}
Thus, the number of parameters is reduced from $5$ to $2$.
As a reference for the optimization, we take the usual canonical choice, i.e., the matrix $\bm{S}=\mathbbm{1}/5$, which gives the same Hamiltonian as obtained by projecting out the $S=2$ subspace obtained from the addition of angular momentum of two spin-1 particles. 
This choice imposes SU(2) symmetry on the Hamiltonian.
After applying the gradient-ascent optimization on $N=30$ sites, we see an improvement in the gap~(\cref{fig:aklt_mps_path}(a)). 
Among the ground states, the SU(2) symmetry is only present at $\lambda=1$, corresponding to the AKLT state~\cite{affleck_quantum_1989}.
Hence, the optimal PH remains different from the canonical ones.
The gains of the optimizations are small in comparison to the random MPS case since the first excited state can switch between different subspaces of the symmetry transformation $O$. 
When the family of PHs is expressed in the symmetric basis of the MPS, the original energy spectrum splits into parts --- each with different value of the quantum number of the symmetry under consideration.
We depict this for the AKLT state in\cref{fig:aklt_mps_path} (b). 
In the subspace containing the ground state, there can be two improvements (i) the spectral gap is larger, since the new first excited state can be one of the higher energy eigenstates of the original Hamiltonian (ii) a non-injective state becomes a unique ground state, which is demonstrated in the next example of GHZ state.

\emph{Adiabatic preparation of GHZ state.---}  
Up to now, we have focused on the case of injective MPS, which are unique ground states of PHs constructed using \cref{eq:local_parent_hamiltonian}.
The same construction in the case of non-injective MPS, leads to Hamiltonian with a degenerate ground space.
The degeneracy can hinder the adiabatic preparation, either when targeting a non-injective MPS as the final state or encountering it as an intermediate state before reaching an injective MPS.
In these situations, it is possible to avoid degeneracies by restricting to a symmetric subspace.
To demonstrate this, we choose the GHZ state which appears during adiabatic evolution between (disordered) cluster state MPS $(\lambda=-1)$ and (ordered) product state $(\lambda=1)$~\cite{wolf_quantum_2006} of spin-$\frac{1}{2}$ particles, given by the following tensors,
\begin{align}
\label{eq:mps_ghz}
A_0 = 
\begin{pmatrix}
    0&0
    \\
    1&1
\end{pmatrix},
A_1 = 
\begin{pmatrix}
    1&\lambda
    \\
    0&0
\end{pmatrix}.
\end{align}

\begin{figure}
\centering
\includegraphics[width=\columnwidth]{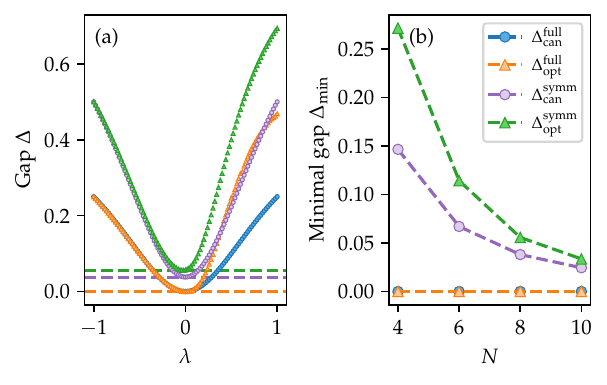}
\caption{(a) Illustration of gaps along the canonical $(\mathrm{can})$ and optimized $(\mathrm{opt})$ adiabatic path for $N=8$ sites sweeping from product state $(\lambda=-1)$ to cluster state $(\lambda=1)$, crossing the GHZ state at $(\lambda=0)$;
$\mathrm{full}$ refers to the full Hilbert space, $\mathrm{symm}$ to the symmetric subspace with the ground state.
(b) Minimum gaps along the path. 
}
\label{fig:ghz_mps_path}
\end{figure}

At $\lambda=0$, the above tensor corresponds to the GHZ state, which is non-injective, and the family of PHs generated by \cref{eq:local_parent_hamiltonian} is 2-fold degenerate.
Thus, in the full Hilbert space, the spectral gap closes as $\lambda$ approaches 0 (c.f. orange curve in panel (a) of \cref{fig:ghz_mps_path}).
To lift this degeneracy, we impose conservation of parity of the $\sigma_x$ operator $P=\prod_{i=1}^N\sigma^x_i$ and translation invariance of the tensors, enforcing a non-zero gap (c.f. purple curve in \cref{fig:ghz_mps_path}).
However, the gap still decays with increasing system size and at sufficiently large $N$, the optimized gaps converge to the gaps of the canonical Hamiltonian $(\bm{S}=\mathbbm{1}/4)$ (c.f. panel (b) in \cref{fig:ghz_mps_path}).

\emph{Conclusion and Outlook.---}
In this work, we optimize the Hamiltonian path for preparing specific TNSs adiabatically. 
We interpolate between the MPS tensors and build the corresponding PHs with maximum gap along the path.
The fixed locality of the PH optimization is a desirable property for experiments.
We demonstrate our algorithm in three paradigmatic cases, random MPS preparation, the AKLT state and the non-injective GHZ state, yielding uniformly better gaps in all cases.
By exploiting symmetries, the minimum gap along the path is further enhanced, even circumventing the appearance of degeneracies in PHs where the tensors become non-injective.
Finally, we formulate our adiabatic preparation method as a differential equation for the optimal parameters, yielding an implicit solution equivalent to pointwise optimization.
With additional information about the model, an analytical solution for the optimal parameters may be possible.
Ultimately, our method provides an avenue to reduce the overall time required for adiabatic tensor network state preparation.

Our method requires access to the spectral gap of the PH, which may be hard to obtain in practice.
Instead of an exact computation, we can replace the gap estimation by alternate methods, which target systems beyond the validity region of DMRG, \textit{e.g.}~higher geometrical dimensions or in the thermodynamic limit.
A similar approach to Refs.~\cite{cruz_preparation_2022,kull_lower_2024} can be used by substituting exact values of the spectral gap with lower bounds derived from semidefinite programming methods~\footnote{in preparation}.
The use of lower bounds instead of exactly computed gaps is a natural next step.

Our research opens up several future questions.
While we demonstrate the method through a linear deformation in the tensors $A^{(i),\sigma}_{\alpha,\beta}(\lambda)$, a natural step is
understanding non-linear deformations and the extension to multi-variable parameter spaces. 
The study of the minimum gap in these different parametrizations can provide an in-depth interpretation of the different phases of TNSs~\cite{chen_classification_2011, schuch_classifying_2011}.
Additionally, going to the multi-parameter regime is expected to provide a better control over the optimization space.
Secondly, for real adiabatic preparation, the fidelity of the final state depends not only on the gap, but also on the speed of traversal~\cite{schiffer_adiabatic_2022,chen_speeding_2022}.
The combined optimization of speed and spectral gap is expected to further enhance the efficiency of adiabatic preparation algorithms. 

\acknowledgements
\emph{Acknowledgements.} ---
We thank Norbert Schuch, Esther Cruz Rico, Benjamin Schiffer, Stefano Polla, Georgios Styliaris, Ilya Kull, Flavio Baccari, and Zhi-Yuan Wei, for useful discussions.
P.E. and J.T. acknowledge the support received by the Dutch National Growth Fund
(NGF), as part of the Quantum Delta NL programme. 
P.E. acknowledges the support received through the NWO-Quantum Technology
programme (Grant No.~NGF.1623.23.006).
K.S.R., J.F.C., and J.T. acknowledge the support received from the European Union's Horizon Europe research and innovation programme through the ERC StG FINE-TEA-SQUAD (Grant No.~101040729). 
This publication is part of the `Quantum Inspire - the Dutch Quantum Computer in the Cloud' project (with project number [NWA.1292.19.194]) of the NWA research program `Research on Routes by Consortia (ORC)', which is funded by the Netherlands Organization for Scientific Research (NWO).
Parts of this work were performed by using the compute
resources from the Academic Leiden Interdisciplinary Cluster Environment (ALICE) provided by Leiden University.

The code used for performing the simulations in this letter will be available upon publishing.

The views and opinions expressed here are solely
those of the authors and do not necessarily reflect those of the funding institutions. Neither
of the funding institutions can be held responsible for them.

\bibliography{references}

\clearpage
\newpage
\onecolumngrid
\crefname{equation}{Eq.}{Eqs.}
\setcounter{section}{0}
\crefname{figure}{Fig.}{Fig.}
\crefname{appendix}{Appendix}{Appendix}
\renewcommand{\thefigure}{S\arabic{figure}}
\renewcommand{\theequation}{S\arabic{equation}}
\makeatletter
\begin{center} 
	{\large \bf Supplementary Materials: Spectral Gap Optimization for Enhanced Adiabatic State Preparation}
\end{center} 

The supplementary materials are devoted to providing detailed derivations in the main text.
In section~\ref{sec:parent_hamiltonian}, we detail the construction of parent Hamiltonians for a given tensor network state.
Next, we state the Weyl's inequality and show that it implies the concavity of the gap as a function of the parameters of the Hamiltonian in section~\ref{sec:weyl_inequality}.
In section~\ref{sec:optimality}, we derive the conditions on the Hamiltonian parameters to achieve the optimal value for the gap.
Additionally, we mention how the conditions can be used in the situation when we have access to one of the parent Hamiltonians, but not the ground state MPS.
In section~\ref{sec:random_mps}, we discuss the method of preparation of random injective MPS and how the parent Hamiltonian is constructed.
In section~\ref{sec:symmetry}, we show how symmetries of the tensor network state can be used to reduce the number of unknown parameters in the Hamiltonian, which helps in scaling the problem to higher system sizes. 
We focus on the case of the AKLT and GHZ states.
Finally, in section~\ref{sec:numerics}, we mention the algorithms and packages used to perform the simulations. Additionally, we also mention the specific values of parameters for the DMRG algorithm.
\section{Construction of the Parent Hamiltonian \label{sec:parent_hamiltonian}}
We consider \textit{normal} tensors whose parent Hamiltonians can be built by blocking the tensors into injective tensors~\cite{cirac_matrix_2021}. 
Normal tensors span the vector space of $D\times D$ matrices upon enough blocking of sites. 
The following equation is a necessary condition for injectivity,
\begin{align}
    d^L\geq D^2,
\end{align}
where $L$ denotes the number of blocked sites.
For a general PEPS, the map from the virtual to physical indices after blocking a region $R$ is of the form
\begin{align}
    \mathcal{T}(B):(\mathbb{C}^D)^{\otimes \vert\partial R\vert}\rightarrow (\mathbb{C}^d)^{\otimes R},
\end{align}
where $B$ denotes the injective tensor formed from blocking tensors $A$ on $L$ sites.
Then, the Hamiltonian is built out of projectors in $(\text{Im}\mathcal{T})^{\perp}$ which will be orthogonal to the tensor network state which lies in $(\text{Im}\mathcal{T})$.
We assume the tensor $A$ become injective after blocking two sites.
The orthogonal subspace $(\text{Im}\mathcal{T})^{\perp}$ has the property
\begin{align}
\left\{ \Tr(XA^{(j),\sigma_{j}}A^{(j+1),\sigma_{j+1}})=0:X\text{ arbitrary}\right\}.
\end{align}
For the above orthogonal subspace on site $j$ and $j+1$ is spanned by vectors $\text{span}\{\ket{\phi_{j,k}}\}_k$,
and $k = 1,...,r_j$ labels the vectors in the basis, $r_j$ being the total number of vectors. 
Then,
\begin{align}
\label{eq:parent_hamiltonian_S_ab}
    h_{j,j+1} = \sum_{\alpha,\beta=1}^{r_j} S_{j,\alpha\beta}\ket{\phi_{j,\alpha}}\bra{\phi_{j,\beta}},
\end{align}
where $S_j\succ 0$ is positive-definite at all sites $j$, ensuring that rest of the spectrum is above zero.
The full Hamiltonian can be written as,
\begin{align}
    \label{eq:eigendecomposition_of_H}
    H(S):=\sum_j\sum_{\alpha,\beta=1}^{r_j}S_{j,\alpha\beta}\ket{\phi_{j,\alpha}}\bra{\phi_{j,\beta}} = \sum_{l=0}^{d^N-1}E_l(S)\ket{E_l(S)}\bra{E_l(S)}=\sum_{l=1}^{d^N-1}E_l\ket{E_l}\bra{E_l},
\end{align}
where we have dropped the $S$ for brevity of notation and the last equality follows from the ground state energy being 0. 
For our purpose, the working form of the parent Hamiltonian in the rest of the text is Eq.~\eqref{eq:parent_hamiltonian_S_ab}.
Owing to the construction, these Hamiltonians are frustration-free because the ground state MPS minimizes the local terms simultaneously~\cite{schuch_peps_2010}.
Furthermore, for the uniqueness of the ground state of the parent Hamiltonian, the initial tensors $A$ are required to be \textit{normal}~\cite{cirac_matrix_2021}. 
Suppose the injectivity length (required blocking length) is $L$, then constructing the parent Hamiltonian after blocking $L+1$ sites ensures that the ground state is unique.
\section{Concavity of spectral gap of parent Hamiltonians}
\label{sec:weyl_inequality}
The parameterized family of parent Hamiltonians of an MPS have a common ground state space $P_G$ which contains the MPS.
To separate the ground state subspace $(P_G)$ from the rest of the spectrum $(P_G^{\perp})$, we apply a unitary transformation as follows,
\begin{equation}
\label{eq:separated_H}
    UH (s)U^{\dagger} = P_G^{\perp}(s)\bigoplus P_G,
\end{equation}
where $P_G^{\perp}$ has dimension $(d^N-1)\cross (d^N-1)$. 
The transformation divides the spectrum into two parts: $P_G$ is independent of $s$, and $P_G^{\perp}$ is dependent.
A convex combination of two such parent Hamiltonians can be written as,
\begin{align}
\label{eq:concavity_P_G_perp}
\begin{split}
    UH(s)U^{\dagger} &= 
    U\left[(1-s)H(0)+s H(1)\right]U^{\dagger}
    \\&=\Big[(1-s)P_G^{\perp}(0)+s P_G^{\perp}(1)\Big] \bigoplus P_G.
\end{split}
\end{align}
Here $H(0)$ and $H(1)$ represent two parent Hamiltonians constructed with a fixed choice of the parameters. 
Comparing Eqs.~\eqref{eq:separated_H} and \eqref{eq:concavity_P_G_perp}, the relation obtained is 
\begin{align}
    P_G^{\perp}(s) = (1-s)P_G^{\perp}(0)+s P_G^{\perp}(1).
\end{align}
This means, that the concavity property of parent Hamiltonians is inherited by the corresponding subspaces $P_G^{\perp}$.
Next, to show the concavity of the gap, we state the Weyl's inequality first.
\begin{lemma} (Weyl's inequality~\cite{weyl_asymptotische_1912})
Let $A$ and $B$ be two Hermitian matrices on the same inner product vector space $V$, and $\mu_i(A)$ and $\mu_i(B)$ for $i=1,...,n$ denote the $i^{\text{th}}$ eigenvalues of $A$ and $B$ respectively in increasing order.
Then, 
    \begin{align}
        \mu_{i+j-1}(A+B)\geq \mu_i(A)+\mu_j(B)\geq\mu_{i+j-n}(A+B).
    \end{align}
\end{lemma}
Denoting the spectrum of $P_G^{\perp}(\lambda)$ by $E_i^{\perp}(\lambda)$ for $i=1,..,d^N-1$, then from Weyl's inequality $(i=j=1)$, we obtain that
\begin{equation}\label{eq:concavity_condition}
E_1^{\perp}(s) \geq (1-s)E_1^{\perp}(0)+s E^{\perp}_1(1).
\end{equation}
Note that the minimum eigenvalue of $P_G^{\perp}$ corresponds to the spectral gap of $H$.
Applying Eq.~\eqref{eq:concavity_condition} to the spectral gap of $H$, we get the relation
\begin{align}
    \Delta(s)\geq(1-s)\Delta(0)+s\Delta(1).
\end{align}
This proves the concavity of the gap of parent Hamiltonians.
\section{Optimal parameter space evolution with $\lambda$}
\label{sec:optimality}
In this section, we derive the requirements for optimality of the gap and the differential equation (5) in the main text.
The optimality condition can be used to find the matrix $\bm{S}_\mathrm{opt}$ for a certain choice of the basis $\{\ket{\phi_i}\}_i$.
In the subsequent equations we will replace the ket notation with the matrix notation of $\Phi$ to represent the kernel basis. 
The Jacobian and Hessian of the gap with respect to the parameter matrix $\bm{S}$ are defined as
\begin{equation}
    \mathcal{\bm{J}}_{\Delta}(\bm{S}):=\frac{\partial\Delta(\Phi(\lambda),\bm{S})}{\partial\bm{S}},\label{eq:def_of_G}
\end{equation}
and
\begin{equation}
     \bm{\mathcal{H}}_{\Delta}(\bm{S})=\frac{\partial^{2}\Delta (\Phi(\lambda),\bm{S})}{\partial\bm{S}^{2}(\lambda)},
 \end{equation}
 respectively.
The optimal gap is reached when $\bm{\mathcal{J}}_{\Delta}(\bm{S}_\mathrm{opt}(\lambda))=0$ and $\bm{\mathcal{H}}_{\Delta}(\bm{S}_\mathrm{opt}(\lambda))\preceq 0$.

\subsection{Requirements for optimality of the gap}
We detail the steps required to obtain the optimality of the gap.
This includes computing the analytic expressions for the Jacobian and the Hessian of the gap with respect to the positive matrix $\bm{S}$.

The positive matrix $\bm{S}$ can be expressed in terms of a Hermitian matrix $\bm{B}$ as follows:
\begin{equation}
    \bm{S}=\frac{e^{-\bm{B}}}{\mathrm{tr}(e^{-\bm{B}})},
\end{equation}
which establishes a bijection between the random choice $\bm{B}$ and positive definite matrix $\bm{S}$. 
Furthermore, it guarantees that $\bm{S}$ is normalized with $\Tr(\bm{S})=1$.
Consider a basis for Hermitian matrices $\{\bm{V}_k\}_k$, then in terms of the basis elements, $\bm{S}$ is
\begin{equation}
    \bm{S}=\sum_k s_k \bm{V}_k\succeq 0.
\end{equation} 
The gap $\Delta(\Phi,\bm{S})$ is fully determined by the local Hamiltonian $h_i$, i.e. by the matrix $\bm{S}$ and the basis for the kernel of the injective map $\Phi:=\{\ket{\phi_\alpha}\}_\alpha$. 

The calculations below will also be applicable for the cases where the Hamiltonian is the restricted to the symmetric subspace.
For simplicity, we retain the full Hilbert space consideration in this section.
If the first excited state is $\ket{\psi_1}$, the spectral gap is 
\begin{equation}
    \Delta(\Phi,\bm{S})=\left\langle \Psi_{1}\right|H(\Phi,\bm{S})\left|\Psi_{1}\right\rangle,
\end{equation}
since the ground state energy of $H$ is zero by construction.
Given the basis $\Phi$, the parent Hamiltonian is
\begin{equation}
    H(\Phi,\bm{S})=\sum_{i,\alpha,\beta}S_{\alpha\beta}\ket{\phi_{i,\alpha}}\bra{\phi_{i,\beta}},
\end{equation}
which leads to the following expression for the gap
\begin{align}
\Delta(\Phi ,\bm{S})
&=
\sum_{i,\alpha,\beta}S_{\alpha\beta}\bra{\Psi_{1}}\ket{\phi_{i,\alpha}} \bra{ \phi_{i,\beta}}\ket{\Psi_{1}}
=\sum_{i,\alpha,\beta}S_{\alpha\beta}\chi_{i,\alpha\beta}
=\sum\limits_i\Tr(\bm{S}\chi^T_i)
=\sum_i\langle \bm{S},\chi_i^T\rangle
\label{eq:gapwithchi}
\end{align}
where $\langle M,N \rangle = \Tr(M^\dagger N)$ denotes the Hilbert-Schmidt inner product of matrices $M$ and $N$ and
We define $\chi:=\sum_i\chi_i$, then the gap has the following dependence
\begin{align}
    \Delta(\Phi,\bm{S}) = \langle\bm{S},\chi^T\rangle.
\end{align}
To compute the Jacobian at fixed value of $\lambda$ (which means the basis $\Phi$ is fixed), we use the Hellmann-Feynman theorem~\cite{guttinger_verhalten_1932} and obtain 
\begin{align}
    \frac{\partial\Delta}{\partial\bm{V}_k} 
    = 
    \expval**{\frac{\partial H(\Phi,\bm{S})}{\partial\bm{V}_k}}{\Psi_1}
    =\sum_{\alpha,\beta}\frac{\partial}{\partial\bm{V}_k}\left(\frac{e^{-\bm{B}}}{\Tr(e^{-\bm{B}})}\right)_{\alpha\beta}\chi_{\alpha\beta}
    =\left\langle\frac{\partial}{\partial\bm{V}_k}\left(\frac{e^{-\bm{B}}}{\Tr(e^{-\bm{B}})}\right),\chi^T\right\rangle.
    \label{eq:DVk_gap}
\end{align}
Using the following definition of matrix derivative, 
\begin{equation}
\partial_{\bm{V}_k}e^{-\bm{B}}=\int_{0}^{1}e^{(x-1)\bm{B}}\bm{V}_ke^{-x \bm{B}}dx,
\end{equation}
we get,
\begin{equation}
\partial_{\bm{V}_k}\Tr (e^{-\bm{B}})=\Tr(\partial_{\bm{V}_k}e^{-\bm{B}})=\Tr(\bm{V}_ke^{-\bm{B}}).
\end{equation}
Using the above two equations
\begin{equation}
\partial_{\bm{V}_{k}}\frac{e^{-\bm{B}}}{\Tr(e^{-\bm{B}})}
=\frac{\int_{0}^{1}e^{(x-1)\bm{B}}\bm{V}_{k}e^{-x\bm{B}}dx}{\Tr(e^{-\bm{B}})}
-\frac{e^{-\bm{B}}\Tr(\bm{V}_{k}e^{-\bm{B}})}{\Tr(e^{-\bm{B}})^{2}}.\label{eq:DVk_expression}
\end{equation}
Substituting Eq.~\eqref{eq:DVk_expression} into Eq.~\eqref{eq:DVk_gap}, 
we obtain 
\begin{equation}
    \partial_{\bm{V}_k}\Delta(\Phi,\bm{S})
    =
    \left\langle\frac{\int_{0}^{1}e^{(x-1)\bm{B}}\bm{V}_{k}e^{-x\bm{B}}dx}{\Tr(e^{-\bm{B}})},\chi^T
    \right\rangle
    -
    \left\langle\frac{e^{-\bm{B}}\Tr(\bm{V}_{k}e^{-\bm{B}})}{\Tr(e^{-\bm{B}})^{2}}
    ,\chi^T
    \right\rangle.
\label{eq:optimalconditionforS}
\end{equation}

Without loss of generality, we work with a basis $\ket{\phi_{i,\alpha }}$ in which $\bm{S}$ and $\bm{B}$ are diagonal.
This means that the local term of the Hamiltonian will be a sum of projectors. 
We make $\bm{S}$ diagonal by taking $\bm{S}_{\alpha\beta}=\bm{S}_{\beta \beta }\delta_{\alpha \beta}$.
Further, in the treatment below, we will take the following matrix basis $E^{\gamma_1\gamma_2}$,
\begin{equation}
    E^{\gamma_1\gamma_2}_{\alpha \beta}=\delta_{\gamma_1\alpha}\delta_{\gamma_2\beta},
\end{equation}
where $\gamma_1,\gamma_2$ parameterize the basis elements, and $E^{\gamma_1\gamma_2}_{\alpha \beta}$ is the $(\alpha,\beta)$ element of $E^{\gamma_1\gamma_2}$.
For $E^{\gamma_1\gamma_2}_{\alpha \beta}$ with $\gamma_1\ne \gamma_2 $, we have $\Tr(E^{\gamma_1\gamma_2}e^{-B})=0$, and thus
\begin{equation}
\partial_{E^{\gamma_1\gamma_2}}\Delta(\Phi,\bm{S})=\frac{1}{\Tr(e^{-\bm{B}})}\frac{e^{-b_{\gamma_2}}-e^{-b_{\gamma_1}}}{b_{\gamma_1}-b_{\gamma_2}}\chi_{\gamma_2\gamma_1}.\label{eq:D_E_alphabeta}
\end{equation}
where $b_{\gamma_i}$ are the diagonal elements of $\bm{B}$.
The derivative with respect to $E^{\gamma\gamma}$ is
\begin{align}
\partial_{E^{\gamma\gamma}}\Delta(\Phi ,\bm{S})&=\sum_{\beta}\left[\frac{E^{\gamma\gamma}e^{-b_{\gamma}}}{\Tr(e^{-\bm{B}})}-\frac{e^{-\bm{B}}e^{-b_{\gamma}}}{\Tr(e^{-\bm{B}})^{2}}\right]_{\beta\beta}\chi_{\beta\beta}\\&=\frac{e^{-b_{\gamma}}}{\Tr(e^{-\bm{B}})}\left[\chi_{\gamma\gamma}-\frac{\sum_{\beta}e^{-b_{\beta}}\chi_{\beta\beta}}{\Tr(e^{-\bm{B}})}\right]\\
&=S_{\gamma \gamma}[\chi_{\gamma \gamma }-\Tr(\bm{S}\chi^T)].\label{eq:D_E_betabeta}
\end{align}

Assuming that the gap is smooth at the optimal point $\bm{S}_\mathrm{opt}$,
all the first-order derivatives will vanish at the optimal point. 
Then \cref{eq:D_E_alphabeta} implies that 
\begin{align}
    \chi_{\mathrm{opt},\alpha\beta}=0,\quad\forall\,\alpha\neq\beta.
\end{align}
This means that the two matrices $\bm{S}_\mathrm{opt}$ and $\chi_\mathrm{opt}$ are simultaneously diagonalizable. 
Here, $\chi_\mathrm{opt}$ denotes the matrix at the optimal $\bm{S}_\mathrm{opt}$.
Further requiring that \cref{eq:D_E_betabeta} is equal to zero, we get that
\begin{align}
    S_{\mathrm{opt},\gamma\gamma}&=0,
    \\
    \text{ or }\nonumber
    \\
\chi_{\mathrm{opt},\gamma\gamma}&=\Tr(\bm{S}_\mathrm{opt}\chi^T_{\mathrm{opt}}),\label{eq:optimal_S_results}
\end{align}
Since one of the above conditions should always hold, there will be an eigenvalue $\chi_{\mathrm{opt},\gamma\gamma}$ of $\chi_{\mathrm{opt}}$ with multiplicity $k$ such that,
\begin{align}
    k = \text{rank}(\bm{S})
\end{align}
and the rest of the eigenvalues will be independent.
Moreover, \cref{eq:gapwithchi} and \cref{eq:optimal_S_results} imply that the value of $\chi_{\mathrm{opt},\gamma\gamma}$ will exactly correspond to the optimal gap
\begin{equation}\Delta_{\mathrm{opt}}=\langle\bm{S}_{\mathrm{opt}},\chi_{\mathrm{opt}}^{T}\rangle=\sum_{\beta}S_{\mathrm{opt},\beta\beta}\chi_{\mathrm{opt},\beta\beta}=\chi_{\mathrm{opt},\gamma\gamma}\mathrm{Tr}(\bm{S}_{\mathrm{opt}})=\chi_{\mathrm{opt},\gamma\gamma},
\end{equation}
where we use that $\bm{S}$ is diagonal and $\Tr(\bm{S}_\mathrm{opt})=1$.
The construction of requirements of optimality applies directly in the case when the ground state MPS is unknown, but a corresponding parent Hamiltonian is given.

More formally, given the guarantee that a given Hamiltonian $H_1 = \sum_ih_i$ is frustration-free, it is possible to define a family of Hamiltonians $H(\bm{Q})$ which will have the same ground state as of $H_1$.
Note that this does not require the knowledge of the ground state itself.
The image of $h_i$ will correspond to the projectors $P_{G^{\perp},i}$ onto the orthogonal space of the ground state which we denote as $G^\perp$.
A general family of Hamiltonians can then be written as $h=\vec{a}\cdot\vec{P}_{G^{\perp}}$, with appropriate normalization constraints on $a$. Here, $\vec{a}$ denotes the vector of strictly positive coefficients that parameterize the family.
Once the family of parent Hamiltonians is defined, the Hamiltonian optimization can be performed in the same manner as for the family of Hamiltonians generated from MPS tensors.
\subsection{Derivation of differential equation (5) in main text}
As discussed in the previous section, if the gap is smooth at the optimal $\bm{S}_\mathrm{opt}$, the Jacobian is zero at that point. Thus, it remains zero during the complete adiabatic path,
\begin{equation}
    \bm{\mathcal{J}}_{\Delta}(\bm{S}_\mathrm{opt}(\lambda))\equiv 0,\,\,\forall\,\lambda.
\end{equation}
We compute the variation of the Jacobian with $\lambda$ as,
 \begin{align}
\frac{d\bm{\mathcal{J}}_{\Delta}(\bm{S}_\mathrm{opt} (\lambda))}{d\lambda}
&=\frac{d}{d\lambda}\left.\left(\frac{\partial\Delta}{\partial\bm{S}}\right)\right|_{\bm{S}=\bm{S}_\mathrm{opt}(\lambda)}
\\&=\left.\left(\frac{\partial^{2}\Delta}{\partial\Phi\partial\bm{S}}\right)\right|_{\bm{S}=\bm{S}_\mathrm{opt}(\lambda)}\frac{\partial\Phi}{\partial\lambda}+\left.\left(\frac{\partial^{2}\Delta}{\partial\bm{S}^2}\right)\right|_{\bm{S}=\bm{S}_\mathrm{opt}(\lambda)}\frac{\partial\bm{S}_\mathrm{opt}}{\partial\lambda}
\\&=\left(\frac{\partial}{\partial\Phi}\bm{\mathcal{J}}_{\Delta}(\bm{S}_\mathrm{opt}(\lambda))\right)\frac{\partial\Phi}{\partial\lambda}+\bm{\mathcal{H}}_{\Delta}(\bm{S}_\mathrm{opt}(\lambda))\frac{\partial\bm{S}_\mathrm{opt}}{\partial\lambda}=0,
\end{align}
which gives the Eq. (5) in the main text.
\section{Preparation of random injective MPS}
\label{sec:random_mps}
We will use the preparation method discussed in \cite{cruz_preparation_2022} and adapt it for the case of translation-invariant MPS.
A non translation-invariant MPS can be written as follows,
\begin{align}
    \ket{\psi(\mathcal{\vec{P}})} = \bigotimes_{j=1,\mathrm{odd}}^N\mathcal{P}_{j+1,j+2}\left(\sum_{i=1}^d\ket{ii}\right)_{j,j+1},
\end{align}
where first the entangled pairs are generated and then the projector maps are applied.
We will consider the case where the physical dimension is the same as the bond dimension $d=D$, and the injective tensor is obtained by blocking two sites, i.e. $d^2=D^2$.
\begin{figure}[h]
    \centering
    \includegraphics[width=0.5\textwidth]{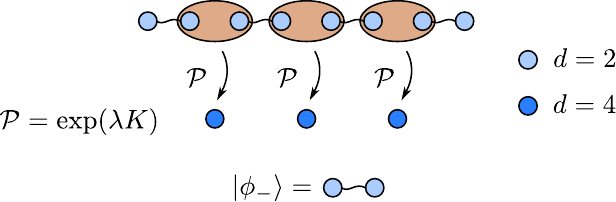}
    \caption{Preparation of random injective MPS: first prepare entangled pairs and then apply positive maps $\mathcal{P}$ on adjacent sites from two consecutive pairs. The map $\mathcal{P}$ is parametrized as the exponential of a random Hermitian matrix, giving a positive matrix $\mathcal{P}$, thus ensuring injectivity.}
    \label{fig:1dpeps}
\end{figure}
The map $\mathcal{P}$ can be further broken down into $\mathcal{P} = QW$, where $Q$ is a positive definite matrix and $W$ is a unitary.
Then, to prepare translation-invariant MPS, we apply the same maps on all sites, $\mathcal{P}_i=\mathcal{P} = e^{\lambda K}$, where $K$ is a random Hermitian matrix.

We will now discuss the specifications of the random MPS, based on which the numerics in the main text were performed.
The first step is to take a product state on all spin-$\frac{1}{2}$'s such as the all-up state $\ket{\uparrow}^{\otimes N}$.
Then we prepare entangled pairs in the $\ket{\phi_-} = \frac{1}{\sqrt{2}}\left(\uparrow\downarrow-\downarrow\uparrow\right)$ state using the unitary $U$.
In the next step we apply maps of the form $\mathcal{P}=e^{\lambda K}$ for random choices of $K$, which are matrices of size $(4\cross 4)$.
Upon blocking two sites, this gives us a new MPS with $D=2$ and $d=4$.
The canonical parent Hamiltonian is chosen as shown in \cite{cruz_preparation_2022}.
The particular form for the local Hamiltonian terms $h_j$ are the following, for odd values of $j$
\begin{align}
h_j &= O_j^{\dagger}\Pi_{j,j+1}O_j
\\
O_j&=\left(U^{-1}\right)_{j,j+1}\left(e^{-\lambda K}\right)_{j+1,j+2}\left(e^{-\lambda K}\right)_{j-1,j}
\\
\Pi_{j,j+1} &= \left(\ket{\downarrow}\bra{\downarrow}\right)_j+\left(\ket{\downarrow}\bra{\downarrow}\right)_{j+1}
\end{align}
and for even values of $j$,
\begin{align}
h_j &= O_j^{\dagger}\Pi_{j-1,j}O_j
\\
O_j&=\left(U^{-1}\right)_{j-1,j}\left(e^{-\lambda K}\right)_{j-2,j-1}\left(e^{-\lambda K}\right)_{j,j+1}
\\
\Pi_{j-1,j} &= \left(\ket{\downarrow}\bra{\downarrow}\right)_{j-1}+\left(\ket{\downarrow}\bra{\downarrow}\right)_j
\end{align}
Thus, the canonical parent Hamiltonian is 5-local in the original spin-$\frac{1}{2}$ particles picture.
Since the final MPS has physical dimension $d=4$, we block two sites to get the injective form.
The injective map $\mathcal{T}$ is 
\begin{align}
    \mathcal{T}(B):(\mathbb{C}^2)^{\otimes 2}\rightarrow (\mathbb{C}^4)^{\otimes 2},
\end{align}
where $Y$and typically, the dimension of $(\mathrm{Im}\mathcal{T})^{\perp}$ is $4^2-2^2 = 12$.
This means that the parent Hamiltonian is $2-$local in the $d=4$ (spin-$\frac{3}{2}$) particle space. 
For constructing the family of parent Hamiltonians, the parametrization is done using a positive definite matrix (c.f. section \ref{sec:parent_hamiltonian}) of size $(12\times 12)$, and thus there are 144 free parameters, which we optimize using a gradient-ascent algorithm.
\section{Optimization of the gap in symmetry sectors}
\label{sec:symmetry}
In this section, we show that the parent Hamiltonian with highest gap inherits the symmetry of the starting tensor network state.
Suppose the state is symmetric under the action of group $G$, i.e., $U_g \ket{\psi(A)}=\ket{\psi(A)}$ $\forall\,g\in G$ if $U_g$ is the unitary representation of the group, and $A$ represents the tensors of the MPS representation of the state. 
Next, we consider a parent Hamiltonian $H$ that does not have this symmetry and apply the unitary transformation corresponding to all elements of the symmetry group of the ground state and average over the Haar measure. 
This gives the symmetry-preserving parent Hamiltonian,
\begin{align}
    H_\mathrm{symm}=\frac{1}{|G|}\sum_{g\in G} U_g H U_g^\dagger=\int_G d\mu(g) U_g H U_g^\dagger, 
\end{align}
where the first expression is for a discrete symmetry group, and the second is for a compact Lie symmetry group with $\mu(g)$ being the normalized Haar measure.
Since $U_g$ is unitary, the individual terms in the sum $U_g H U_g^\dagger$ have the same spectrum as $H$, implying that $\Delta(U_g H U_g^\dagger)=\Delta(H)$. From the concavity property of the spectral gap of parent Hamiltonians (Section~\ref{sec:weyl_inequality}),
\begin{align}
    \Delta (H_\mathrm{symm})\geq \frac{1}{|G|}\sum_{g\in G}\Delta(U_{g}HU_{g}^{\dagger})=\int_G d\mu(g)\Delta(U_{g}HU_{g}^{\dagger})=\Delta(H).
\end{align}
which means that the gap of symmetry-preserving Hamiltonian is at least as large as the original Hamiltonian $H$. 
This means that optimizing over $H_\mathrm{symm}$ leads to the optimal parent Hamiltonian for the TN under consideration.
Additionally, the family of symmetric Hamiltonians, $H_\mathrm{symm}$, have a block diagonal structure with each of the blocks representing a different value of the quantum number corresponding to the symmetry.
In the discussion below, we fix the quantum number to the value which it has in the ground state, which means the new Hamiltonian will be given by one of the blocks of $H_{\mathrm{symm}}$.
Moreover, the gap in the subspace becomes a smooth function of the matrix $\bm{S}$. 
For achieving larger gaps and a smooth function for the gap, we consider the parent Hamiltonians restricted to the sector satisfying the symmetries of the ground state.

For the AKLT model, the total $S^z_\mathrm{tot}=\sum_{i=1}^N S^z_i$ is conserved~\cite{affleck_rigorous_1987}, i.e., $[S^z_\mathrm{tot},H(\lambda)]=0$, and the ground state lies in the sector $S^z_\mathrm{tot}=0$. 
The AKLT state also has $\mathbb{Z}_2$ symmetry
\begin{equation}
    Q=\prod_{i=1}^N S_i^x R,
\end{equation}
where $R$ reverses the state, i.e., $R\ket{i_1,i_2,...,i_N}=\ket{i_N,i_N-1,...,i_1}$. 
Note that $Q$ commutes with $(S^{z}_\mathrm{tot})^2$ but not with $S^z_\mathrm{tot}$. 
Therefore, we can split the complete Hilbert space by conserving $S^z_\mathrm{tot}$ and $T$ separately. 
The subspace with $S^z_{\mathrm{tot}}=0$ can be further divided into parts with $R=1$ and $R=-1$. 
The ground state lies in the subspace with $T=1$, $S^z_\mathrm{tot}=0$, and $Q=1$.
Similarly, the GHZ state has the parity symmetry $P=\prod_{i=1}^N\sigma^x_i$ and the reversal symmetry $R$ and it lies in the subspace with $T=1$, $R=1$, and $P=1$.

\subsection{Optimized technique for AKLT state preparation}
\begin{figure}
    \centering
    \includegraphics[width=14.5cm]{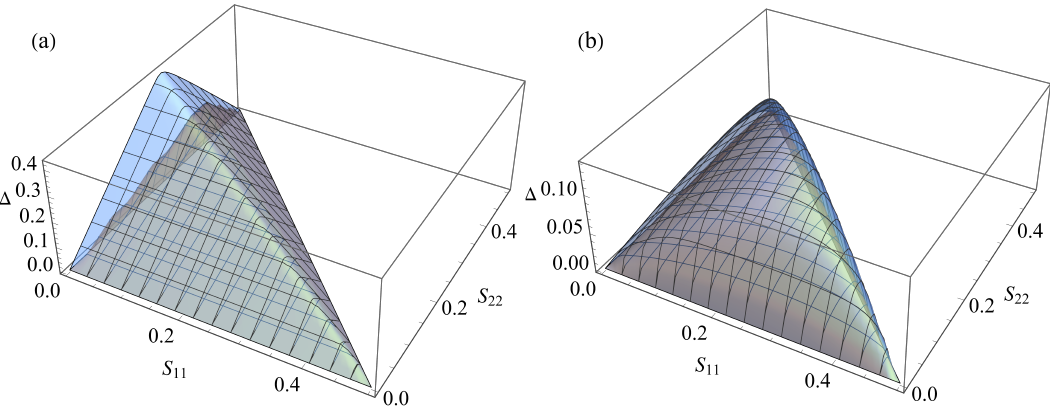}
    \caption{AKLT state preparation $N=8$ sites:
    The gaps $\Delta$ in the full Hilbert space and the symmetry subspace containing the ground state --- as functions of $S_{11}$ and $S_{22}$. (a) $\lambda=0.1$ and (b) $\lambda=0.7$. Note: The gap in the full Hilbert space is not smooth with $\bm{S}$ at the optimal $\bm{S}_\mathrm{opt}$ for $\lambda=0.1$. 
    }
    \label{fig:gapN33D}
\end{figure}

\begin{figure}
    \centering
\includegraphics[width=16cm]{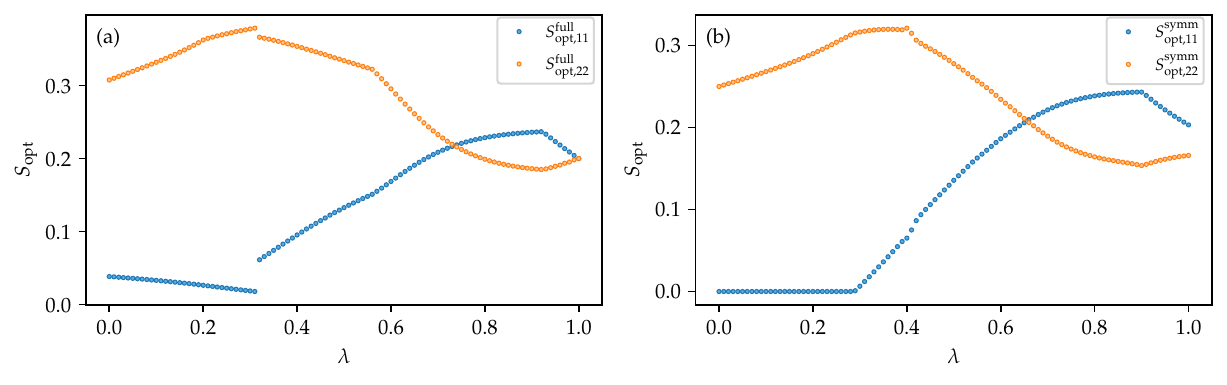}
    \caption{
    AKLT state preparation $N=8$ sites: The optimal $S_{\mathrm{opt},11}$ and $S_{\mathrm{opt},22}$ in (a) the full Hilbert space and (b) the symmetry subspace containing the ground state.
    $\bm{S}_{\mathrm{opt}}$ is discontinuous in the full Hilbert space, but continuous in the symmetry subspace.}
\label{fig:AKLT_optimized_subspace}
\end{figure}

We construct the parent Hamiltonian for the adiabatic preparation
of the AKLT state. 
As discussed in the main text, we choose a linear deformation of a qutrit MPS ($\ket{+1} $, $\ket{0}$,
and $\ket{-1}$) interpolating from a product state of qutrits ($\ket{0}^{\otimes N}$) to the AKLT state,
\begin{equation}
A^{+1}(\lambda)=\left(\begin{array}{cc}
0 & -\lambda\\
0 & 0
\end{array}\right),\;A^{0}(\lambda)=\left(\begin{array}{cc}
\frac{1}{\sqrt{2}} & 0\\
0 & -\frac{\lambda}{\sqrt{2}}
\end{array}\right),\;A^{-1}(\lambda)=\left(\begin{array}{cc}
0 & 0\\
\lambda & 0
\end{array}\right).\label{eq:AKLT_family_state}
\end{equation}
The state at $\lambda=1$ corresponds to the AKLT MPS.
For this family of states (\ref{eq:AKLT_family_state}), the map from the virtual to physical indices becomes injective after blocking two
sites
\begin{equation}
\mathcal{P}_{\mathrm{inj}}=\left(\begin{array}{c}
A^{+1}A^{+1}\\
A^{+1}A^{0}\\
A^{+1}A^{-1}\\
A^{0}A^{+1}\\
A^{0}A^{0}\\
A^{0}A^{-1}\\
A^{-1}A^{+1}\\
A^{-1}A^{0}\\
A^{-1}A^{-1}
\end{array}\right)\overset{\mathrm{reshape}}{=}\left(\begin{array}{cccc}
0 & 0 & 0 & 0\\
0 & \frac{\lambda^{2}}{\sqrt{2}} & 0 & 0\\
-\lambda^{2} & 0 & 0 & 0\\
0 & -\frac{\lambda}{\sqrt{2}} & 0 & 0\\
\frac{1}{2} & 0 & 0 & \frac{\lambda^{2}}{2}\\
0 & 0 & -\frac{\lambda^{2}}{\sqrt{2}} & 0\\
0 & 0 & 0 & -\lambda^{2}\\
0 & 0 & \frac{\lambda}{\sqrt{2}} & 0\\
0 & 0 & 0 & 0
\end{array}\right).
\end{equation}
For accessing the projectors for the parent Hamiltonian, we choose the following basis $\{\ket{\phi_i}\}_i$ for
$(\text{Im}\mathcal{P}_\mathrm{inj})^{\perp}$,
\begin{align}
\left|\phi_{1}(\lambda)\right\rangle  & =\left|+1,+1\right\rangle \\
\left|\phi_{2}(\lambda)\right\rangle  & =\frac{\left|+1,0\right\rangle +\lambda\left|0,+1\right\rangle }{\sqrt{1+\lambda^{2}}}\\
\left|\phi_{3}(\lambda)\right\rangle  & =\frac{\left|+1,-1\right\rangle +2\lambda^{2}\left|0,0\right\rangle +\lambda^{2}\left|-1,+1\right\rangle}{\sqrt{1+5\lambda^{4}}}\\
\left|\phi_{4}(\lambda)\right\rangle  & =\frac{\left|0,-1\right\rangle +\lambda\left|-1,0\right\rangle }{\sqrt{1+\lambda^{2}}}\\
\left|\phi_{5}(\lambda)\right\rangle  & =\left|-1,-1\right\rangle .
\end{align}
Writing the ket vectors as columns of a matrix $\Phi$ we get,
\begin{align}\label{eq:aklt_Psi}
    \Phi = 
    \begin{pmatrix}
        1&0&0&0&0&0&0&0&0
        \\
        0&\frac{1}{\sqrt{1+\lambda^2}}&0&\frac{\lambda}{\sqrt{1+\lambda^2}}&0&0&0&0&0
        \\
        0&0&{\frac{1}{\sqrt{1+5\lambda^{4}}}}&0&\frac{2\lambda^2}{\sqrt{1+5\lambda^4}}&0&\frac{\lambda^2}{\sqrt{1+5\lambda^4}}&0&0
        \\
        0&0&0&0&0&\frac{1}{\sqrt{1+\lambda^2}}&0&\frac{\lambda}{\sqrt{1+\lambda^2}}&0
        \\
        0&0&0&0&0&0&0&0&1
    \end{pmatrix}^T.
\end{align}
A natural choice for the parent Hamiltonian can be obtained by taking equal contribution (diagonal elements) of all projectors,
\begin{equation}
\bm{S}=\mathbbm{1}_5/5,
\end{equation}
At $\lambda=1$, this choice corresponds to the standard AKLT Hamiltonian~\cite{affleck_rigorous_1987},
\begin{align}
h_{i,i+1} & =\left[\frac{1}{2}\vec{S}_i\cdot\vec{S}_{i+1}+\frac{1}{6}(\vec{S}_i\cdot\vec{S}_{i+1})^{2}+\frac{1}{3}\right].\label{eq:hAKLT_normal}
\end{align}
The MPS state satisfies three symmetries characterized by the translation operator $T$, the total $S^z_\mathrm{tot}=\sum_{i=1}^N S^z_i$, and the operator $Q=\prod_{i=1}^NS_i^x R$ where $R$ corresponds to the reflection operator. 
For the parent Hamiltonian with these symmetries, the matrix $\bm{S}$ has the form, 
\begin{align}
    \bm{S}=\left(\begin{array}{ccccc}
S_{11}\\
 & S_{22}\\
 &  & S_{33}\\
 &  &  & S_{22}\\
 &  &  &  & S_{11}
\end{array}\right),
\end{align}
where the off-diagonal elements are all zero and the diagonal elements are symmetric along the anti-diagonal. 
The local term is explicitly
\begin{equation}
    h(\lambda)=S_{11} h_{11}(\lambda)+S_{22} h_{22}(\lambda)+S_{33} h_{33}(\lambda),
\end{equation}
where $h_{11}(\lambda)$, $h_{22}(\lambda)$, and $h_{33}(\lambda)$ can be written as 
\begin{align}
    (h_{11}(\lambda))_{i,i+1}&=\frac{2}{9}\mathbbm{1}
    +\frac{1}{6}\tau^3_{i+1}
    -\frac{1}{6 \sqrt{3}}\tau^8_{i+1}
    +\frac{1}{6}\tau^3_i
    +\frac{1}{4}\tau^3_i\tau^3_{i+1}
    +\frac{1}{4 \sqrt{3}}\tau^3_i\tau^8_{i+1}
    -\frac{1}{6 \sqrt{3}}\tau^8_i+\frac{1}{4 \sqrt{3}}\tau^8_i\tau^3_{i+1}
    +\frac{5}{12} \tau^8_i\tau^8_{i+1}
    \\
    (h_{22}(\lambda))_{i,i+1}&=\frac{2}{9}\mathbbm{1}-
    \frac{1}{6(\lambda^2+1)}\tau^3_{i+1}
    +\frac{\left(2 \lambda ^2-1\right)}{6 \sqrt{3} \left(\lambda ^2+1\right)}\tau^8_{i+1}
    +\frac{\lambda}{2 \left(\lambda ^2+1\right)}\tau^1_i\tau^1_{i+1}
    +\frac{\lambda}{2 \left(\lambda ^2+1\right)}\tau^2_i\tau^2_{i+1}
    -\frac{\lambda ^2}{6 \left(\lambda ^2+1\right)}\tau^3_i
    \nonumber
    \\&-\frac{1}{4}\tau^3_i\tau^3_{i+1}
    -\frac{\left(\lambda ^2-3\right)}{4 \sqrt{3} \left(\lambda ^2+1\right)}\tau^3_i\tau^8_{i+1}
    +\frac{\lambda}{2 \left(\lambda ^2+1\right)}\tau^6_i\tau^6_{i+1}
    +\frac{\lambda}{2 \left(\lambda ^2+1\right)}\tau^7_i\tau^7_{i+1}
    -\frac{\left(\lambda ^2-2\right)}{6 \sqrt{3} \left(\lambda ^2+1\right)}\tau^8_i+
    \\&+\frac{\left(3 \lambda ^2-1\right)}{4 \sqrt{3} \left(\lambda ^2+1\right)}\tau^8_i\tau^3_{i+1}
    -\frac{1}{12}\tau^8_i\tau^8_{i+1}\nonumber\\
    (h_{33}(\lambda))_{i,i+1}&=\frac{1}{9}\mathbbm{1}-\frac{\lambda ^4}{2 \left(5 \lambda ^4+1\right)}\tau^3_{i+1}
    +\frac{\left(5 \lambda ^4-2\right)}{6 \sqrt{3} \left(5 \lambda ^4+1\right)}\tau^8_{i+1}
    +\frac{\lambda ^2}{5 \lambda ^4+1}\tau^1_i\tau^6_{i+1}
    +\frac{\lambda ^2}{5 \lambda ^4+1}\tau^2_i\tau^7_{i+1}
    +\frac{\left(1-4 \lambda ^4\right)}{6 \left(5 \lambda ^4+1\right)}\tau^3_i
    \nonumber
    \\&+\frac{\lambda ^4}{5 \lambda ^4+1}\tau^3_i\tau^3_{i+1}
    -\frac{\left(2 \lambda ^4+1\right)}{2 \sqrt{3} \left(5 \lambda ^4+1\right)}\tau^3_i\tau^8_{i+1}
    +\frac{\lambda ^2}{2 \left(5 \lambda ^4+1\right)}\tau^4_i\tau^4_{i+1}
    +\frac{\lambda ^2}{2 \left(5 \lambda ^4+1\right)}\tau^5_i\tau^5_{i+1}
    +\frac{\lambda ^4}{5 \lambda ^4+1}\tau^6_i\tau^1_{i+1}
    \\&+\frac{\lambda ^4}{5 \lambda ^4+1} \tau^7_i\tau^2_{i+1}
    +\frac{\left(2 \lambda ^4+1\right)}{6 \sqrt{3} \left(5 \lambda ^4+1\right)}\tau^8_i
    -\frac{\sqrt{3} \lambda ^4}{2 \left(5 \lambda ^4+1\right)}\tau^8_i\tau^3_{i+1}
    +\frac{\left(\lambda ^4-1\right)}{2 \left(15 \lambda ^4+3\right)}\tau^8_i\tau^8_{i+1}\nonumber,
\end{align}
where $\tau_i$ for $i=1,...,8$ denote the Gell-Mann matrices.
\begin{table}
\centering
\begin{tabular}{|c|c|c|c|c|c|c|c|c|c|c|c|}
\hline 
$N$ & 2 & 3 & 4 & 5 & 6 & 7 & 8 & 9 & 10 & 11 & 12 \tabularnewline
\hline 
\hline 
$d^N$ & 9 & 27 & 81 & 243 & 729 & 2187 & 6561 & 19683 & 59049 & 177147 & 531441
\tabularnewline
\hline $d_\mathrm{g}$ & 2 & 3 & 6 & 10 & 22 & 42 & 98 & 216 & 532 & 1288 & 3326
\tabularnewline
\hline 
\end{tabular}
\caption{Dimension of the whole Hilbert space $d^N$ and the ground-state subspace $d_\mathrm{g}$ with different $N$ for AKLT model. \label{tab:Dimension-of-the}}
\end{table}
In Table~\ref{tab:Dimension-of-the}, the dimension of the ground state subspace are listed in comparison with the dimension of the Hilbert space of the system.
We show the dimensions of the ground-state subspace and the whole Hilbert space in Table~\ref{tab:Dimension-of-the}
In Fig.~\ref{fig:gapN33D}, we plot the gap in the whole Hilbert space and in the ground state subspace as a function of $S_{11}$ and $S_{22}$ for 8 sites.
In Fig.~\ref{fig:AKLT_optimized_subspace}, we plot ${S_{\mathrm{opt},11}}$ and ${S_{\mathrm{opt},22}}$ for 8 sites.
We observe that, in the full Hilbert space, $\bm{S}_\mathrm{opt}$ is full rank and the gap can be a non-smooth function of $\bm{S}$. 
On the other hand, in the symmetry subspace, $\bm{S}_\mathrm{opt}$ is not full rank and the gap is smooth with $\bm{S}$.

\FloatBarrier
\subsection{Optimized technique for GHZ state preparation}
In this section, we analyze the transition from product state of spin-1/2's to the cluster state using the MPS,
\begin{align}
\label{eq:GHZexampleMPS}
        A^0 = 
        \begin{pmatrix}
            0&0
            \\
            1&1
        \end{pmatrix},
        A^1 = 
        \begin{pmatrix}
            1&\lambda
            \\
            0&0
        \end{pmatrix}
\end{align}
At $\lambda=0$, the GHZ state is formed, which is an example of a non-injective MPS.
Due to the degeneracy in the family of parent Hamiltonians of the GHZ state, the spectral gap closes at $\lambda=0$~\cite{wolf_quantum_2006}.
In this section, we discuss the symmetries present in the above tensor to circumvent this closing of the gap.
 
For achieving injectivity of the MPS (for $\lambda\neq 0$) in Eq.~\eqref{eq:GHZexampleMPS}, we block three sites, and choose the following basis for $(\text{Im}\mathcal{P}_\mathrm{inj})^{\perp}$,
\begin{align}
    \left|\phi_{1}(\lambda)\right\rangle&=\frac{\lambda}{\sqrt{\lambda^{2}+1}}\ket{000}-\frac{1}{\sqrt{\lambda^{2}+1}}\ket{010}\\
    \left|\phi_{2}(\lambda)\right\rangle&=\frac{1}{\sqrt{2}}\ket{001}-\frac{1}{\sqrt{2}}\ket{011}\\
    \left|\phi_{3}(\lambda)\right\rangle&=-\frac{1}{\sqrt{2}}\ket{100}+\frac{1}{\sqrt{2}}\ket{110}\\
    \left|\phi_{4}(\lambda)\right\rangle&=-\frac{1}{\sqrt{\lambda^{2}+1}}\ket{101}+\frac{\lambda}{\sqrt{\lambda^{2}+1}}\ket{111}.
\end{align}
Following the previous procedure (as for AKLT state), we rewrite the basis as the columns of matrix $\Phi$,
\begin{equation}
\Phi=\left(\begin{array}{cccccccc}
\frac{\lambda}{\sqrt{\lambda^{2}+1}} & 0 & -\frac{1}{\sqrt{\lambda^{2}+1}} & 0 & 0 & 0 & 0 & 0\\
0 & \frac{1}{\sqrt{2}} & 0 & -\frac{1}{\sqrt{2}} & 0 & 0 & 0 & 0\\
0 & 0 & 0 & 0 & -\frac{1}{\sqrt{2}} & 0 & \frac{1}{\sqrt{2}} & 0\\
0 & 0 & 0 & 0 & 0 & -\frac{1}{\sqrt{\lambda^{2}+1}} & 0 & \frac{\lambda}{\sqrt{\lambda^{2}+1}}
\end{array}\right)^T.
\end{equation}

For $S_{11}=1/4$, the local term of the parent Hamiltonian is \begin{align}
      h_i(\lambda)&=\frac{1}{4}\Phi_i \Phi_i^\dagger=\frac{1}{8}-\frac{(1+\lambda)^{2}}{16\left(1+\lambda^{2}\right)}\sigma_{i}^{x}-\frac{1-\lambda^{2}}{16\left(1+\lambda^{2}\right)}\sigma_{i}^{z}(\sigma_{i-1}^{z}+\sigma_{i+1}^{z})+\frac{(1-\lambda)^{2}}{16\left(1+\lambda^{2}\right)}\sigma_{i-1}^{z}\sigma_{i}^{x}\sigma_{i+1}^{z},
\end{align}
which is a combination of the transverse field Ising model Hamiltonian and the cluster state Hamiltonian (three-body term)~\cite{wolf_quantum_2006}.
The MPS Eq.~\eqref{eq:GHZexampleMPS} is symmetric under translation operator $T$, the reversal operator $R$, and the operator $P=\prod_{i=1}^N\sigma^x_i$. 
By conserving the expectation values of these operators, the Hilbert space can be split into parts with fixed quantum numbers. 
The dimension of the subspace containing the GHZ state $(d_\mathrm{g})$ is given in Table~\ref{tab:Dimension-of-the_GHZ} in comparison with the size of the Hilbert space.

Next, we construct the parent Hamiltonian with respect to these symmetries. 
The parity operator in the local basis is expressed as 
\begin{equation}
    \bm{P}=\left(\begin{array}{cccc}
 &  &  & 1\\
 &  & 1\\
 & 1\\
1
\end{array}\right). \end{equation}
For the parent Hamiltonian to be symmetric with respect to $\bm{P}$, the $\bm{S}$ matrix satisfies $[\bm{S},\bm{P}]=0$, giving the following simplified form for $\bm{S}$,
\begin{equation}
\label{eq:GHZ_S_mat}
    \bm{S}=\left(\begin{array}{cccc}
S_{11} & S_{12} & S_{13} & S_{14}\\
S_{12} & S_{22} & S_{23} & S_{13}\\
S_{13} & S_{23} & S_{22} & S_{12}\\
S_{14} & S_{13} & S_{12} & S_{11}
\end{array}\right).
\end{equation} 
The matrix is symmetric along both the diagonal and the anti-diagonal.
Since the tensor $A(\lambda)$ by Eq.~\eqref{eq:GHZexampleMPS} is real, i.e., unchanged under complex conjugation. This allows to construct real parent Hamiltonian, and thus we restrict Eq.~\eqref{eq:GHZ_S_mat} to a real matrix.
Now we will analytically illustrate the construction of parent Hamiltonian in the symmetric subspace. 
For simplicity, we restrict the discussion here to the case when $\bm{S}$ is a diagonal matrix, however in the main text, the optimized spectral gap are obtained by taking the general form for $\bm{S}$ as obtained in Eq.~\eqref{eq:GHZ_S_mat}.
Imposing that $\bm{S}$ is diagonal, we get the following form
\begin{equation}
    \bm{S}=\left(\begin{array}{cccc}
S_{11}\\
 & \frac{1}{2}-S_{11}\\
 &  & \frac{1}{2}-S_{11}\\
 &  &  & S_{11}
\end{array}\right),
\end{equation} 
and the corresponding local term is, 
\begin{equation}
    h(\lambda)=S_{11}h_{11}(\lambda)+h_{22}(\lambda),
\end{equation}
where $h_{11}(\lambda)$, $h_{22}(\lambda)$ can be written as
\begin{align}
\left(h_{11}\right)_{i-1,i,i+1}&=\frac{1}{2}\sigma^z_{i-1}\sigma^z_{i+1}+\frac{(\lambda-1)^{2}}{4\left(\lambda^{2}+1\right)}\sigma^x_i-\frac{(\lambda+1)^{2}}{4\left(\lambda^{2}+1\right)}\sigma^z_{i-1}\sigma^x_i\sigma^z_{i+1}+\frac{\left(\lambda^{2}-1\right)}{4\left(\lambda^{2}+1\right)}\left(\sigma^z_{i-1}\sigma^z_i+\sigma^z_i\sigma^z_{i+1}\right)
\\
\left(h_{22}\right)_{i-1,i,i+1}&=\frac{1}{8}\left(\mathbbm{1}-\sigma^x_i-\sigma^z_{i-1}\sigma^z_{i+1}+\sigma^z_{i-1}\sigma^x_i\sigma^z_{i+1}\right).
\end{align}

\begin{table}
\centering
\begin{tabular}{|c|c|c|c|c|c|c|c|c|c|c|c|c|}
\hline 
$N$  & 3 & 4 & 5 & 6 & 7 & 8 & 9 & 10 & 11 & 12 &13&14\tabularnewline
\hline 
\hline 
$d^N$  & 8 & 16 & 32 & 64 & 128 & 256 & 512 & 1024 & 2048 & 4096 & 8192 & 16384
\tabularnewline
\hline 
$d_\mathrm{g}$ & 2 & 4 & 4 & 8 & 9 & 18 & 23 & 44 & 63 & 122 & 190 & 362
\tabularnewline
\hline 
\end{tabular}
\caption{Dimension of the whole Hilbert space $d^N$ and the ground-state subspace $d_\mathrm{g}$ for GHZ model. \label{tab:Dimension-of-the_GHZ}}
\end{table}

\begin{figure}
    \centering
    \includegraphics[width=16cm]{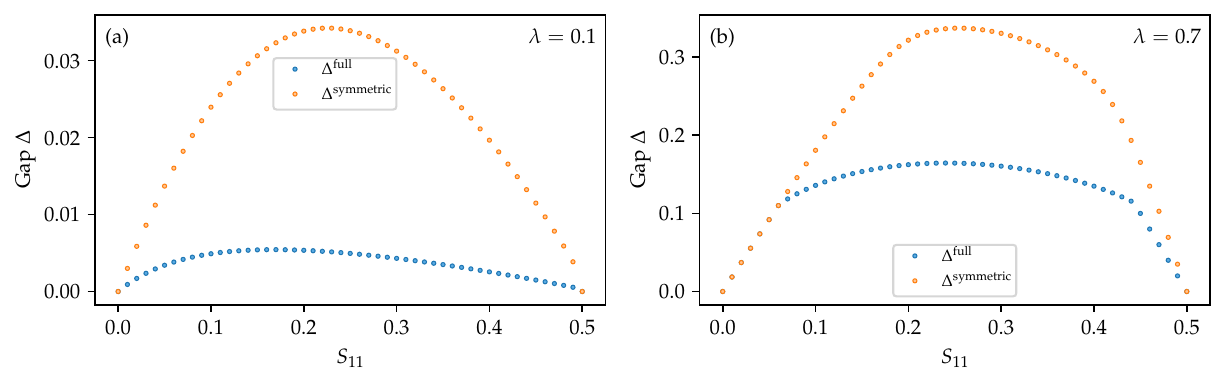}
    \caption{The gaps $\Delta$ in the full Hilbert space and the symmetry subspace as functions of $S_{11}$ for preparing GHZ state with $N=10$. The parameter $\lambda$ is set as (a) $\lambda=0.1$ and (b) $\lambda=0.7$.}
    \label{fig:GHZgapasS11}
\end{figure}

We show the gaps in the whole Hilbert space and in the ground-state subspace as functions of $S_{11}$ in Fig.~\ref{fig:GHZgapasS11}. 
We set the parameter $\lambda=0.1$ in Fig.~\ref{fig:GHZgapasS11}(a) and $\lambda=0.7$ in Fig.~\ref{fig:GHZgapasS11}(b). 
The gap in the whole Hilbert space becomes non-smooth at some particular $S_{11}$ indicating an excited-state phase transition.

\begin{figure}
    \centering
\includegraphics[width=16cm]{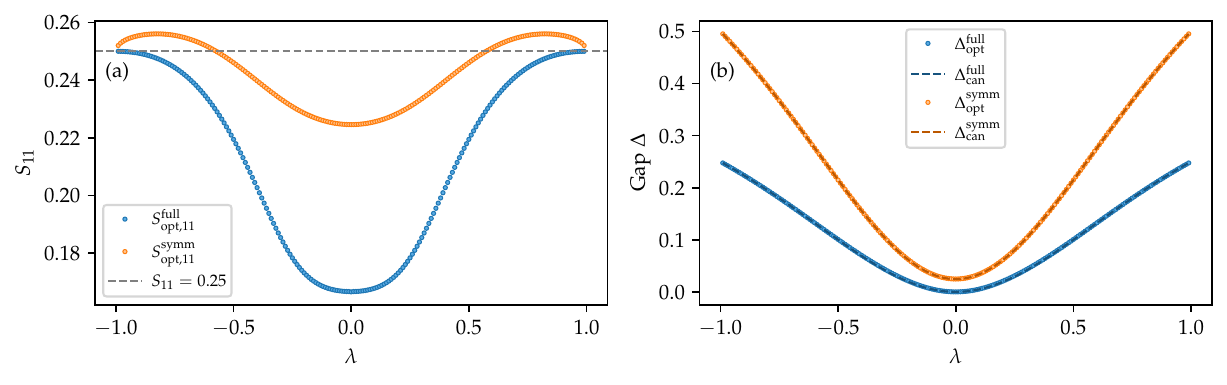}
    \caption{Adiabatic path for the GHZ example $(\lambda=0)$ with $N=10$ sites subject to the constraint that $\bm{S}$ is diagonal: (a) The optimal $S_{\mathrm{opt},11}$ in the full Hilbert space and symmetry subspace. (b) Comparison of $\Delta_{\mathrm{can}}$ (canonical) and $\Delta_{\mathrm{opt}}$ (optimized) in the full Hilbert space and symmetry subspace.}
\label{fig:GHZ_optimized_subspace}
\end{figure}

We further show numerical results of the optimal local term $S_{\mathrm{opt},11}$ and the optimal gap $\Delta_\mathrm{opt}$ in the whole Hilbert space and the ground-state subspace with $N=10$ in Fig.~\ref{fig:GHZ_optimized_subspace}.  
In this example, the optimal gap with the local term $\bm{S}_\mathrm{opt}$ is only slightly larger than that of the maximally mixed local term with $\bm{S}=\mathbbm{1}_4/4$. 
\newpage
\section{Details of Numerics in the main text}
\label{sec:numerics}
The tensor network numerics were mainly performed using Tenpy~\cite{hauschild_efficient_2018}.
Our code also contains the implementation for preparing random non translation-invariant injective MPS, and the corresponding parent Hamiltonians, which was performed using quimb \cite{gray_quimb_2018}.
In all cases, we first prepare the injective maps $\mathcal{P}_{\mathrm{inj}}$ from the tensors, and then use a basis of the kernel $(\mathrm{Im}\mathcal{P}_{\mathrm{inj}})^{\perp}$ for constructing the parent Hamiltonian.

We will discuss the specifications of various parameters used to obtain the Figures 2, 3 and 4 in the main text.

Figure 2: We constructed the MPS as discussed in Section \ref{sec:random_mps} of the supplementary material.
The initial state is a product of $N=24$ spin-$\frac{1}{2}$'s and then the entangled pairs are prepared in the $\ket{\phi_-}$ state.
Next, we apply the positive map $\mathcal{P}$ as shown in section \ref{sec:random_mps} to obtain finally a system of physical dimension $d=4$ (or spin-$\frac{3}{2}$).
The Hermitian matrices $K$ were chosen randomly and the Figure 2 in the main text shows the enhancement in the gap for various choices of the seed values.
The maximum bond dimension used for obtaining the first excited state is set to $D=80$, a reasonable choice given the runtime constraints of our algorithm. 
The DMRG algorithm has to be executed multiple times for each gradient ascent step, as well as for different points along the adiabatic interpolation.
If longer computation times are acceptable, there is no principle hurdle in going to higher bond dimensions.
The maximum energy error and maximum entropy error were $10^{-2}$ and $10^{-8}$ respectively.
The parameter for minimum eigenvalue after SVD is chosen to be $10^{-10}$.

Figure 3(a): We constructed the MPS as discussed in using the tensor in Eq.~7 of the main text, obtaining the AKLT state for $\lambda = 1$.
The simulation is performed on $N=30$ spin-1 sites $(d=3)$.
The maximum energy error and maximum entropy error are $10^{-4}$ and $10^{-8}$ respectively.
The parameter for minimum eigenvalue after SVD is chosen to be $10^{-10}$.

Figure 3(b): For the symmetric subspace optimization on $N=8$ sites we use exact diagonalization to obtain the gap in the symmetry sectors.

Figure 4: We constructed the MPS using the tensor in Eq.~9 of the main text, obtaining the GHZ state for $\lambda = 0$.
Since we require the symmetry restrictions to obtain a non-zero gap (because of non-injectivity of the GHZ state), we use exact diagonalization to perform the simulations on $N=8$ spin-$\frac{1}{2}$ sites $(d=2)$.

All maximizations are performed using the BFGS (Broyden–Fletcher–Goldfarb–Shanno) algorithm~\cite{fletcher_practical_2000}. 
The gradients are computed exactly using the Hellmann-Feynman theorem.
\end{document}